\documentclass[
    ,final            
  ]
  {aipproc}

\layoutstyle{8x11double}

\begin{document}

\title{Spectroscopic and Photometric Observations of Kepler Asteroseismic Targets}


\author{J.\ Molenda--\.Zakowicz}{
  address={Astronomical Institute, University of Wroc\l aw, ul.\ Kopernika 11, 51-622, Wroc\l aw, Poland}
}

\author{M.\ Jerzykiewicz}{
  address={Astronomical Institute, University of Wroc\l aw, ul.\ Kopernika 11, 51-622, Wroc\l aw, Poland}
}

\author{G.\ Kopacki}{
  address={Astronomical Institute, University of Wroc\l aw, ul.\ Kopernika 11, 51-622, Wroc\l aw, Poland}
}

\author{A.\ Frasca}{
  address={Catania Astrophysical Observatory, via S.\ Sofia 78, 95123, Italy}
}

\author{G. Catanzaro}{
  address={Catania Astrophysical Observatory, via S.\ Sofia 78, 95123, Italy}
}

\author{D.W.\ Latham}{
  address={Harvard-Smithsonian Center for Astrophysics, 60 Garden Street, Cambridge, MA 02138, USA}
}

\author{E.\ Niemczura}{
  address={Astronomical Institute, University of Wroc\l aw, ul.\ Kopernika 11, 51-622, Wroc\l aw, Poland}
}

\author{A.\ Narwid}{
  address={Astronomical Institute, University of Wroc\l aw, ul.\ Kopernika 11, 51-622, Wroc\l aw, Poland}
}

\author{M.\ St\k{e}\'slicki}{
  address={Astronomical Institute, University of Wroc\l aw, ul.\ Kopernika 11, 51-622, Wroc\l aw, Poland}
}

\author{T.\ Arentoft}{
  address={Department of Physics and Astronomy, Danish AsteroSeismology Center (DASC), Aarhus University, Ny Munkegade 120, Bldg. 1520, 8000 \AA rhus, Denmark}
}

\author{J.\ Kubat}{
  address={Astronomical Institute AV CR -- Ondrejov, Fricova 298, 25165  Ondrejov, Czech Republic}
}

\author{D.\ Drobek}{
  address={Astronomical Institute, University of Wroc\l aw, ul.\ Kopernika 11, 51-622, Wroc\l aw, Poland}
}

\author{W. Dimitrow}{
  address={Astronomical Observatory of A.\ Mickiewicz University, ul.\ S\l oneczna 36, 60-286 Pozna\'n, Poland}
}

\begin{abstract}
We summarize our ground-based program of spectroscopic and photometric observations 
of the asteroseismic targets of the Kepler space telescope. We have already determined atmospheric 
parameters, projected velocity of rotation, and radial velocity of 62 Kepler asteroseismic 
targets and 33 other stars in the Kepler field of view. 
We discovered six single-lined and two double-lined spectroscopic 
binaries, we determined the interstellar reddening for 29 
stars in the Kepler field of view, and discovered three $\delta$ Sct, two $\gamma$ Dor and 
14 other variable stars in the field of NGC\,6866.
\end{abstract}

\maketitle

\section{Ground-based observations}

Our program of ground-based observations of stars selected for Kepler asteroseismic targets 
has been started in 2005 and is continued since then. We collect high-resolution echelle 
spectra at the Catania Astrophysical Observatory (Italy, 91-cm telescope, observer: JM\.Z), the 
Harvard-Smithsonian Center for Astrophysics (USA, two 1.5-m telescopes and the 6-m MMT
telescope, observer: DWL), the Nordic Optical Telescope (Spain, 2.5-m telescope, observer: TA), 
the Ondrejov Observatory (Czech Republic, 2-m telescope, observers: EN and JK), and the Pozna\'n University 
Observatory in Borowiec (Poland, a single telescope with two 0.5-m mirrors, observer: WD). 
The multicolor CCD and photoelectric data are collected at the Wroc\l aw University Observatory 
in Bia\l k\'ow (Poland, 0.6-m telescope, observers: GK, AN, MS and JM\.Z) and the Catania 
Astrophysical Observatory (Italy, 91-cm telescope, observer: JM\.Z).

We aim at the determination of atmospheric parameters of the program stars, i.e., the 
effective temperature, $T_{\rm eff}$, surface gravity, $\log g$, and metallicity, $\rm 
[Fe/H]$, as well as the projected rotational velocity, $v\sin i$, and radial velocity, 
$v_r$. We use this last parameter to discover and study spectroscopic binaries in the Kepler field
of view. A search for new variable stars in the Kepler field of view is our separate
study.

\section{Results}
\subsection{Atmospheric parameters}
We measured $T_{\rm eff}$, $\log g$, $\rm [Fe/H]$, $v\sin i$, and $v_r$ for 62 Kepler 
asteroseismic targets and 33 other stars in the satellite's field of view (see 
\cite{Molenda2007}, \cite{Molenda2008}, and \cite{Catanzaro2009}). Most of our program 
stars are stars of spectral type F, G or K and the luminosity classes V, IV or III.
However, our sample includes also around 20 A-type stars, a handful of B-type stars, and
one O-type star, HIP\,92637, as well as three subdwarfs, HIP\,92775, HIP\,94704 and 
HIP\,99267, and one supergiant, HIP\,97439.

We find that all but one of the stars selected for Kepler asteroseismic targets have 
solar metallicity or are slightly metal-deficient. The only exception is HIP\,92775, 
sdF8 (see \cite{Molenda2008}); the other two subdwarfs are not Kepler program 
stars.

\subsection{Projected rotational velocity}

The projected velocity of rotation of the F, G and K stars observed by us is low, typically
below 5 km/s (see \cite{Molenda2007} and \cite{Molenda2008}); $v\sin i$ of the 
early-type stars observed by us is significantly higher.

We note, however, that several of our early-type program stars, namely, 
HIP\,93522, A7, HIP\,93941, B2, and HIP\,96762, B9, have $v\sin i$ below 10 km/s (see 
\cite{Catanzaro2009}). This makes them very important and rare targets for an 
asteroseismic study of early-type stars because for stars rotating so slowly 
it is possible to perform an unambiguous identification of the modes of pulsation;
for stars rotating with $v\sin i \gg 10$ km/s the asteroseismic analysis becomes 
difficult and may be inconclusive.

\subsection{Radial velocity}
In \cite{Catanzaro2009} we report a discovery 
of two double-lined spectroscopic binaries, SB2, HIP\,96299 and HIP\,98551.
The former of these stars, 
HIP\,96299, has been discovered to be an eclipsing binary with a period of 10.0486 days by 
\cite{Hartman1994}; the latter, HIP\,98551, is not known to show eclipses. Since for SB2 eclipsing 
binaries it is possible to derive precise masses of the components, which then enter
evolutionary and asteroseismic models as one of the fundamental parameters, both 
stars need further study in order to obtain their radial-velocity curves and find the orbital 
solutions. For HIP98551, time-series photometry would be needed to find out
whether the system is eclipsing.

We have also discovered six single-lined spectroscopic
binaries, SB1, namely, HIP\,94734 and HIP\,94743 discovered by \cite{Molenda2007}, 
HIP\,92132 and HIP\,97513 discovered by \cite{Molenda2008}, and HIP\,96277 and HIP\,97582 
discovered by \cite{Catanzaro2009}. Since for SB1 stars it is possible to calculate 
the systems' mass function, this parameter can be used for estimating the magnitude 
and color indices of the secondary component of the system, and calculating the duplicity 
corrections. Neglecting this corrections may lead to incorrect determination of 
luminosity and effective temperature, and eventually to a wrong asteroseismic model.

\begin{figure}
\caption{The $b-y$ {\it vs.} $\beta$ relation of stars in the Kepler field of view (a 
         figure originally published in \cite{Molenda2009a}). The distances, $r$ [pc], used 
         in coding the symbols were obtained from the revised Hipparcos parallaxes 
         \cite{vanLeeuwen2007}. The solid line represents the standard relation 
         valid for unreddened F0--G2 stars of luminosity classes III--V, the 
         long-dashed line, the standard relation for A7--F0 stars adopted from 
         \cite{Crawford1975} and \cite{Crawford1979}, respectively. The short-dashed 
         lines show by how much an unreddened F-type star may deviate from the standard 
         relation because of ``cosmic scatter'' (see \cite{Crawford1975}). The arrow 
         is the reddening vector $E(b-y) = 0.074$ mag.
}
\includegraphics[height=.3\textheight]{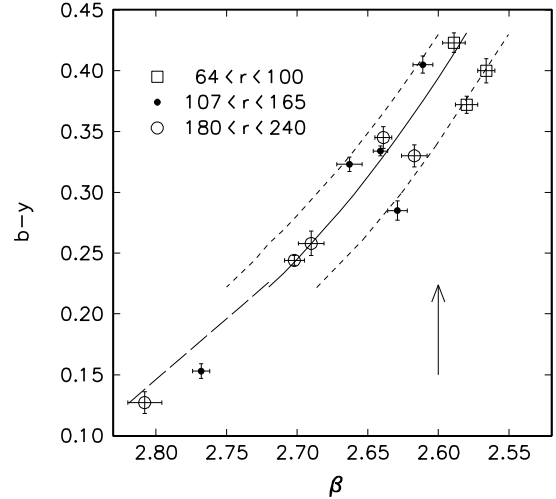}
\end{figure}

\begin{figure}
\caption{The $B-V$ {\it vs.} spectral type diagram of the program stars
(a figure originally published in \cite{Molenda2009a}).
The solid line represents the intrinsic relation for luminosity class V,
and the dashed line, for luminosity class III \cite{Lang1992}. The distances, $r$ [pc],
used in coding the symbols, were obtained from the revised Hipparcos
parallaxes \cite{vanLeeuwen2007}. The two subdwarfs from our sample and two leftmost
luminosity III stars are joined with dotted lines.
The arrow is the reddening vector $E(B-V) = 0.2$ mag.
}
\includegraphics[height=.22\textheight]{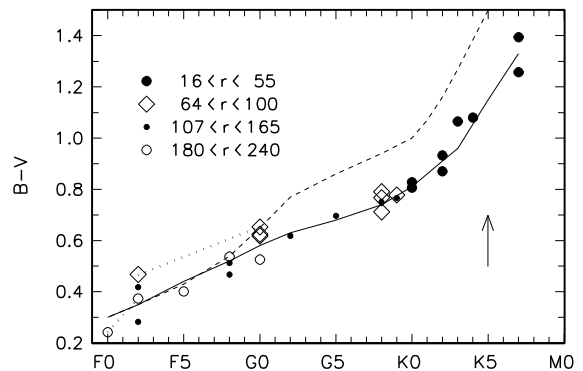}
\end{figure}

\subsection{Interstellar reddening}
In \cite{Molenda2009a}, we report deriving the interstellar reddening for 
29 stars in the Kepler field of view. Having plotted the program stars in several
photometric diagrams we conclude that these stars are not reddened. They do not 
deviate either from the standard relation between $b-y$ and $\beta$ given by 
\cite{Crawford1975} and \cite{Crawford1979} for A7--G2 III--V stars (see Fig.\ 1) or
from the intrinsic relation between $B-V$ and spectral type given by \cite{Lang1992}
for stars of the luminosity class V and III (see Fig.\ 2).
This result is in a disagreement with the information that can be found 
in the Kepler KIC-10 catalogue which gives $E(B-V)$ ranging from 0.01 to 0.06 mag for nine 
of our program stars.

\begin{figure}
\caption{The spectroscopic effective temperature $T_{\rm eff}(1E)$ derived with the use
         of the ELODIE archive, which is an on-line database of high-resolution stellar spectra
         (see http://atlas.obs-hp.fr/elodie), plotted as a function
         of $T_{\rm eff}$ derived from the $(B-V)$ index for Population I stars (points), 
         and $T_{\rm eff}$ derived from the $(B-V)$ and $\delta(0.6)$ indices for the 
         subdwarfs (encircled points). The solid line has 
         unit slope and zero intercept; the short-dashed line runs 311\,K below it.
         (The figure has been originally published in \cite{Molenda2009a}.)}
\includegraphics[height=.23\textheight]{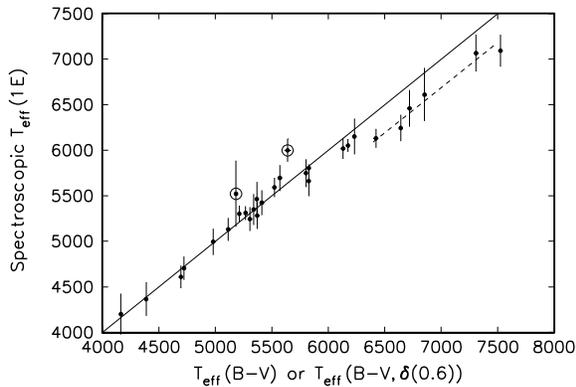}
\end{figure}

Another result of our photometric study of Kepler asteroseismic targets 
is finding a discrepancy between effective temperatures derived for stars hotter 
than 6000\,K by means of spectroscopic and photometric methods (see Fig.\ 3). The 
spectroscopic temperatures of these stars are around 300\,K lower than $T_{\rm eff}$
derived from photometric indices. The reason for this discrepancy is not clear and will
be studied by us in more detail in near future.

\subsection{Variable stars in open clusters}

The open cluster NGC\,6866 is one of four open cluster in Kepler field of view. It
has been selected for a target in an asteroseismic campaign in summer 2009 in the frame of the 
activities of the Kepler Asteroseismic Science Consortium (KASC) Working 
Group\#2.\footnote{Kepler Asteroseismic Science Consortium (KASC) is a group of collaborating 
scientists and/or institutions established to accomplish the activities of the Kepler 
Asteroseismic Investigation (KAI), represented by Ronald Gilliland
(see \url{http://astro.phys.au.dk/KASC}).}

Between April 27 and July 21 2007, we carried out photometric CCD observations of this cluster 
with the aim of a search for variable stars. We used the $BVI_C$ filters of the 
Johnson-Kron-Cousins $UBV(RI)_{\rm C}$ photometric 
system and collected around 470 CCD frames in each filter during 14 nights.

We discovered 19 variable stars of different types (see \cite{Molenda2009b}).
Three of them we classified as $\delta$ Sct, two, as $\gamma$ Dor, four, as W\,UMa, two, as 
ellipsoidal variables, one, as an eclipsing binary and seven, as irregular variables.
All five pulsating variables are very 
probable members of the cluster. This makes them promising asteroseismic targets because
for stars belonging to a cluster it is reasonable to assume the same age and 
metallicity, leaving the radii and masses as the only free parameters in asteroseismic modeling.
The eclipsing binary is definitely not a 
cluster member. Consequently, it is not possible to use this star to measure the age
and distance to NGC\,6866 with the method used by, e.g., \cite{Grundahl2008} and \cite{Meibom2009}. 

Having discovered $\gamma$ Dor stars in NGC\,6866, in \cite{Molenda2009b} we discuss the 
properties of open clusters 
in which pulsators of this type occur. We show that there is no relation between the age 
or metallicity of the cluster and the number of $\gamma$ Dor stars (see Fig.\ 4). In this
way we show that 
the persisting belief that $\gamma$ Dor stars occur only in open clusters which 
are young is unfounded.

\begin{figure}
  \includegraphics[height=.33\textheight]{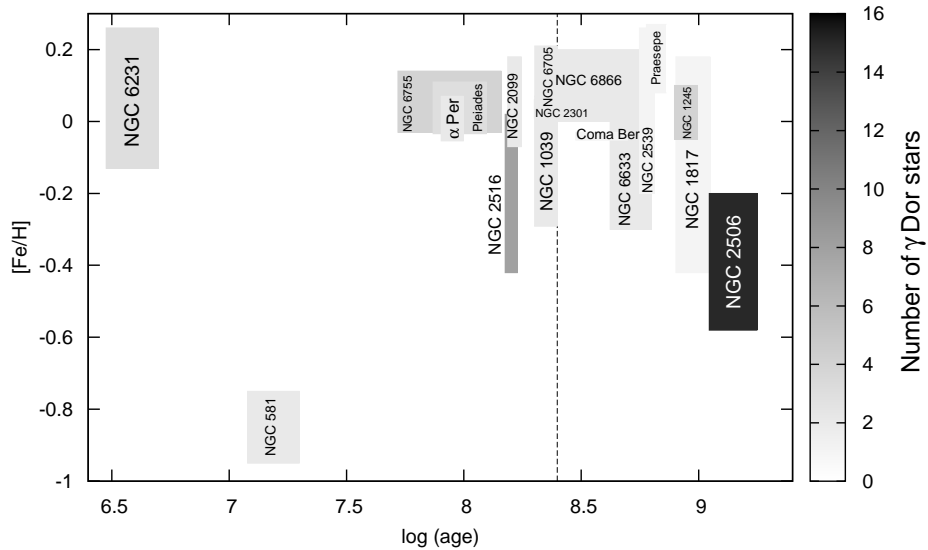}
  \caption{$\gamma$ Dor stars in open clusters coded with shades of gray.
           The vertical line at 250 Myr 
           indicates a suspected upper limit of the age of an open cluster that 
           can host $\gamma$ Dor stars (see \cite{Krisciunas1999}).
           (The figure has been originally published in \cite{Molenda2009a}.)}
\end{figure}

\section{Future work}

\subsection{Constraining parameters of Kepler targets}

Deriving parameters of Kepler targets by means of spectroscopic and photometric
ground-based observations is our primary goal. This task is being realized
at the observatories already mentioned, as well as at 
the Telescopio Nazionale Galileo, TNG, (La Palma, Spain) 
where 19 targets from the KASC Proposal No.\ 30 (see \url{http://astro.phys.au.dk/KASC}) 
have been scheduled for the AOT20/09B service observations (proposal TAC\_71, 
``Spectral characterization of Kepler asteroseismic targets'', P.I.: G.\ Catanzaro). 

In our study we will focus on a detailed analysis of the stars' metallicity, the parameter
which is crucial for asteroseismic analysis of Kepler data. This has been 
demonstrated by \cite{Stello2009} who showed that 
the uncertainty in metallicity dominates the uncertainty in the stellar radius.

We plan also a detailed study of selected program stars with the aim of
solving the discrepancy between the effective temperatures of hot Kepler targets 
determined from spectroscopy and from photometry.

\subsection{Binary stars}

Binary stars require a large amount of observing time. First, they need to be confirmed
as binaries, then, the orbital period must be determined. Depending on the particular target, 
the availability of observing time and the weather conditions, such a task can take several days, weeks or
months. Therefore, we selected a handful of most promising targets which are scheduled for 
observing at several observatories.
We plan to merge all the data we collect and confirm (or reject) the suspected spectroscopic
binarity of these stars, and compute orbital elements for the confirmed ones.

\subsection{Asteroseismic modeling}

We plan to compute evolutionary and asteroseismic models for selected Kepler asteroseismic
targets using the ASTEC and ADIPLS codes (see \cite{JC-D2008a} and \cite{JC-D2008b}). As the 
input of the modeling procedure, 
we will use the atmospheric parameters derived by us in 
\cite{Catanzaro2009}, \cite{Molenda2007}, \cite{Molenda2008} and \cite{Molenda2009a}. 
When possible, we will put additional constraints on the models, e.g., by limiting the possible 
range of masses of eclipsing SB2 stars using the results from their orbital solution.

\section{Data availability}

The echelle spectroscopic observations collected by us at the Catania Astrophysical 
Observatory are available as the HELAS\footnote{HELAS, the European Helio- and 
Asteroseismology Network, has been funded by the European Commission, as a Co-ordination 
Action under th EU's Sixth Framework Programme (FP6) (see \url{http://www.helas-eu.org}}
deliverables at the Wroc\l aw HELAS webpage at \url{http://helas.astro.uni.wroc.pl/deliverables.php}.

In the future, we plan to include in our Internet archive the data collected at other observatories.

\begin{theacknowledgments}
  This work was supported by MNiSW grant N203 014 31/2650.
\end{theacknowledgments}

\end{document}